\pdfoutput=1
\newcommand*{\ATLASLATEXPATH}{}
 
\documentclass[PAPER, atlasdraft=false, texlive=2016, UKenglish, cernpreprint]{\ATLASLATEXPATH atlasdoc}
 
\usepackage[full]{\ATLASLATEXPATH atlaspackage}
\usepackage{\ATLASLATEXPATH atlasbiblatex}
 
\usepackage{\ATLASLATEXPATH atlasphysics}
 
\graphicspath{{logos/}{figures/}}
 
\usepackage{ANA-STDM-2017-26-PAPER-defs}

\AtlasTitle{Evidence for electroweak production of two jets in association with a $Z\gamma$ pair in $pp$ collisions at $\sqrt{s} = 13~\tev$ with the ATLAS detector}
 
\AtlasAbstract{
Evidence for electroweak production of two jets in association with a $Z\gamma$ pair in $\sqrt{s} = 13~\tev$ proton--proton collisions at the Large Hadron Collider is presented. The analysis uses data collected by the ATLAS detector in $2015$ and $2016$ that corresponds to an integrated luminosity of $36.1~\mathrm{fb}^{-1}$. Events that contain a $Z$ boson candidate decaying leptonically into either $e^{+}e^{-}$ or $\mu^{+}\mu^{-}$, a photon, and two jets are selected. The electroweak component is measured with observed and expected significances of $4.1$ standard deviations.
The fiducial cross-section for electroweak production is measured to be $\sigma_{Z\gamma jj\mathrm{-EW}} = 7.8 \; \pm 2.0  \; \mathrm{fb}$, in good agreement with the Standard Model prediction.
}
 
\AtlasJournal{Phys.\ Lett.\ B.}
\AtlasJournalRef{Phys. Lett. B 803 (2020) 135341}
\AtlasDOI{10.1016/j.physletb.2020.135341}
 
\author{The ATLAS Collaboration}
 
\AtlasRefCode{STDM-2017-26}
\PreprintIdNumber{CERN-EP-2019-206}

\AtlasCoverSupportingNote{Measurement of the electroweak Zgammajj cross-section in pp collisions at sqrt(s)=13 TeV with the ATLAS Detector - Supporting Note}{https://cds.cern.ch/record/2304839}
 
\AtlasCoverCommentsDeadline{October 2nd 2019}
 
\AtlasCoverAnalysisTeam{Olympia Dartsi, Lucia Di Ciaccio, Louis Helary, Iro Koletsou, Bruno Lenzi, Shu Li, Narei Lorenzo Martinez, Emmanuel Sauvan, Ben Smart, Yee Chinn Yap}

\AtlasCoverEdBoardMember{Sabine Lammers}
\AtlasCoverEdBoardMember{Juan Terron(chair)}
\AtlasCoverEdBoardMember{Heberth Torres}
 
\AtlasCoverEgroupEditors{atlas-STDM-2017-26-editors@cern.ch}
 
\AtlasCoverEgroupEdBoard{atlas-STDM-2017-26-editorial-board@cern.ch}

\hypersetup{pdftitle={ATLAS document},pdfauthor={The ATLAS Collaboration}}
 
\begin{document}
\sloppy
 
\maketitle

\section{Introduction}
 
At the Large Hadron Collider (LHC), the electroweak (EW) sector of the Standard Model (SM) can be probed by studying the self-coupling of vector bosons, which is precisely predicted through the $\text{SU}(2)_{\mathrm{L}} \times \text{U}(1)_{\mathrm{Y}}$ gauge symmetry.
Vector-boson scattering (VBS), $VV \to VV$ with $V= W/Z/\gamma$, is one of the most interesting processes to look at to study such effects, as it is sensitive to both the triple and the quartic gauge boson couplings, where physics beyond the SM can significantly alter the SM predictions~\cite{Eboli:2003nq,Eboli:2006wa,Degrande:2012wf}.
 
In hadron collisions, VBS events are characterised by the presence of two bosons and two jets, $VV jj$, that are created in a purely electroweak process~\cite{Cahn:1986zv}, through the interaction of bosons that have been radiated from the initial-state quarks. For such events, the scattered quarks are not colour-connected and little hadronic activity is expected in the gap between the two jets. These events are also characterised by a large invariant mass of the dijet system and a large separation of the two jets in rapidity, while the decay products of the bosons are typically produced in the central region.
 
The $VV jj$ final states, however, are produced mainly through a combination of strong and electroweak interactions in $pp$ collisions. The former are of order $\alpha_\text{S}^{2}\alpha_{\mathrm{EW}}^{2}$ at the Born level, where $\alpha_\text{S}$ is the strong coupling constant and $\alpha_{\mathrm{EW}}$ is the electroweak coupling constant. In the following, such processes are referred to as QCD-induced backgrounds, \zgjjqcd. The production of $VVjj$ events through pure $\alpha_{\mathrm{EW}}^{4}$ interactions at the Born level are more rare; they are referred to as electroweak processes, \zgjjew, in the following. It is not possible to study VBS diagrams independently from other electroweak processes as only the ensemble is gauge invariant~\cite{Accomando:2006mc}. There is also interference between the SM electroweak and QCD-induced processes. Some example Feynman diagrams for these processes are shown in Figure~\ref{fig:FeynmanDiagrams}.
 
Among all the possible electroweak production modes, only the $W^{\pm}W^{\pm}jj$ and $WZjj$ production processes have been observed~\cite{Aaboud:2019nmv,Aaboud:2018ddq,Sirunyan:2017ret,Sirunyan:2019ksz}. Evidence for the
$Z\gamma jj$ channel has been reported~\cite{Khachatryan:2017jub}, while limits on electroweak cross-sections have been reported for the $ZZjj$ channel~\cite{Sirunyan:2017fvv}, the $W\gamma jj$ channel~\cite{Khachatryan:2016vif}, and the $VV$ channel~\cite{Aad:2019xxo,Sirunyan:2019der} where at most one of the two bosons decays to two jets.
 
The $Z\gamma jj$ channel~\cite{Aaboud:2017pds} is particularly interesting to study since it allows the measurement of the neutral quartic gauge couplings, as for the $ZZ$ final state but with a larger expected cross-section. The $Z\gamma jj$ cross-sections have been computed at next-to-leading order (NLO) in $\alpha_\text{S}$ for both the QCD-induced backgrounds~\cite{Campanario:2014wga} and the electroweak signal~\cite{Campanario:2017ffz}. The NLO QCD corrections are not included in the following for either the electroweak signal or the QCD-induced background, as they were not evaluated for the phase space used in this paper.
 
This Letter reports the evidence for, and the most precise measurements to date of, electroweak \zgjj production, where the $Z$ boson decays leptonically into either $e^{+}e^{-}$ or $\mu^{+}\mu^{-}$, exploiting the data collected with the ATLAS detector in $2015$ and $2016$ at a centre-of-mass energy of $\sqrt{s} = 13~\TeV$, and corresponding to an integrated luminosity of $36.1~\ifb$.
 
\begin{figure}
\centering
\subfloat[]{\includegraphics[width=0.35\columnwidth]{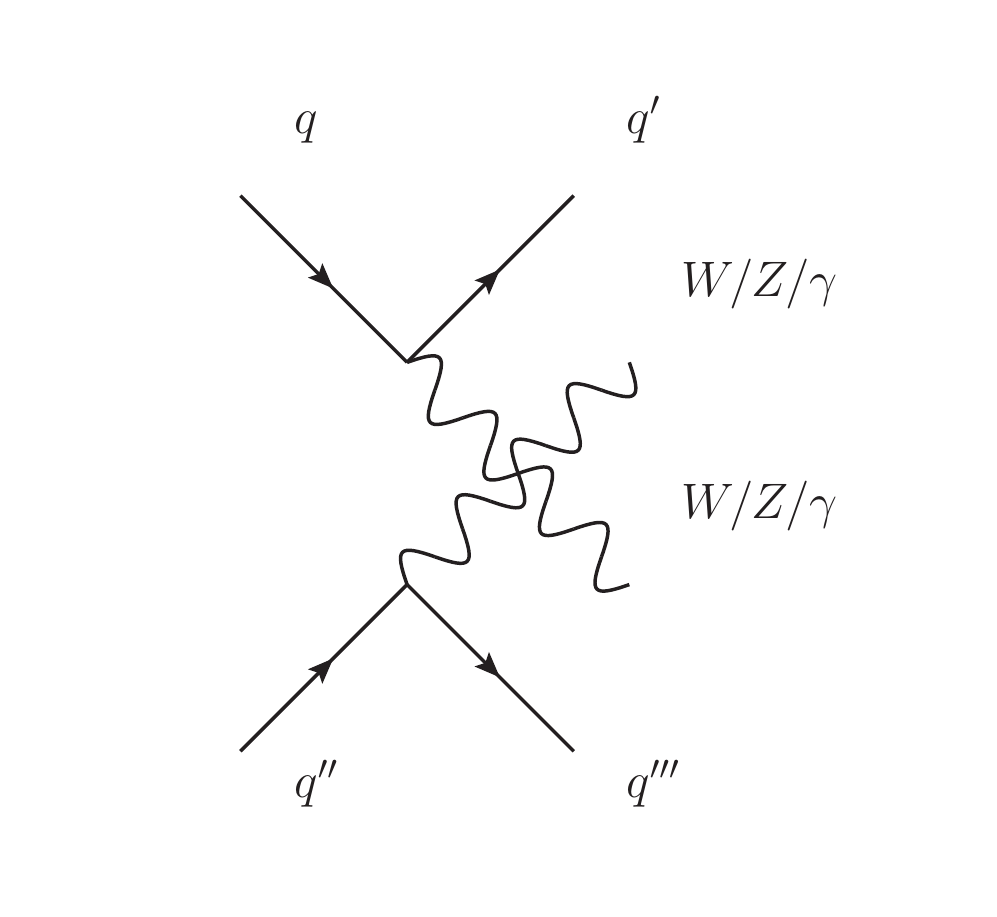}}
\subfloat[]{\includegraphics[width=0.35\columnwidth]{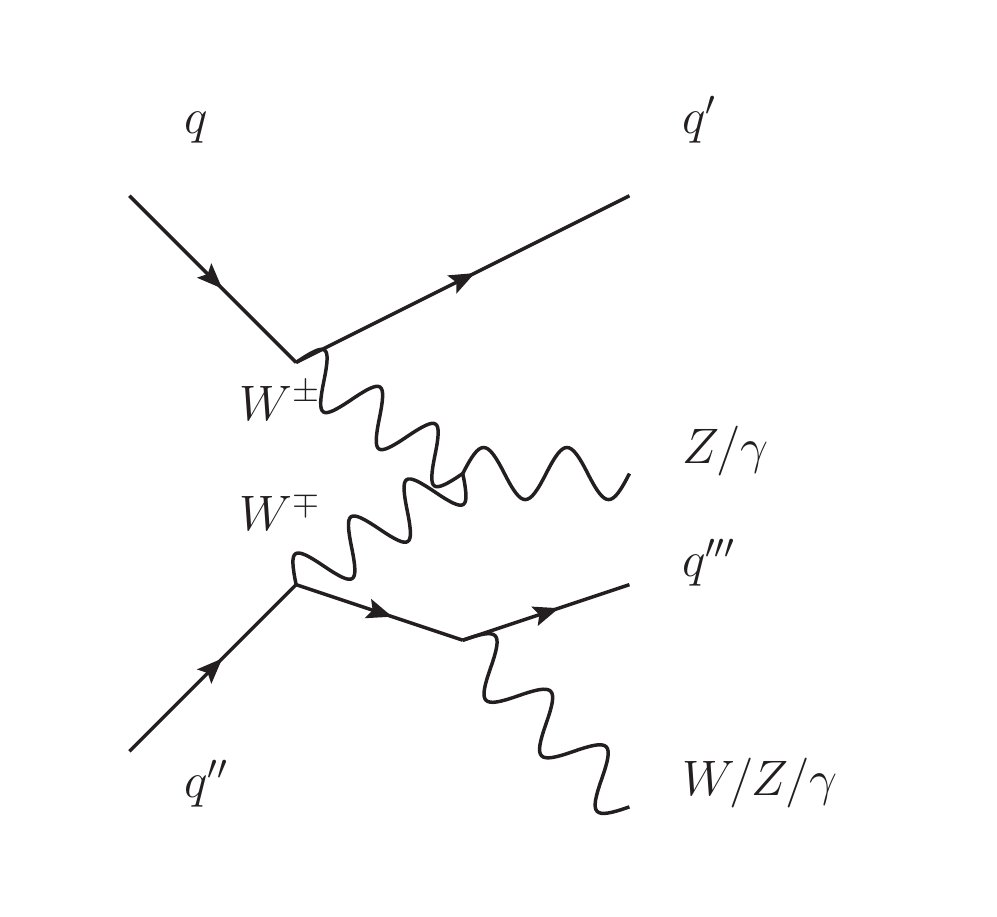}}
\subfloat[]{\includegraphics[width=0.35\columnwidth]{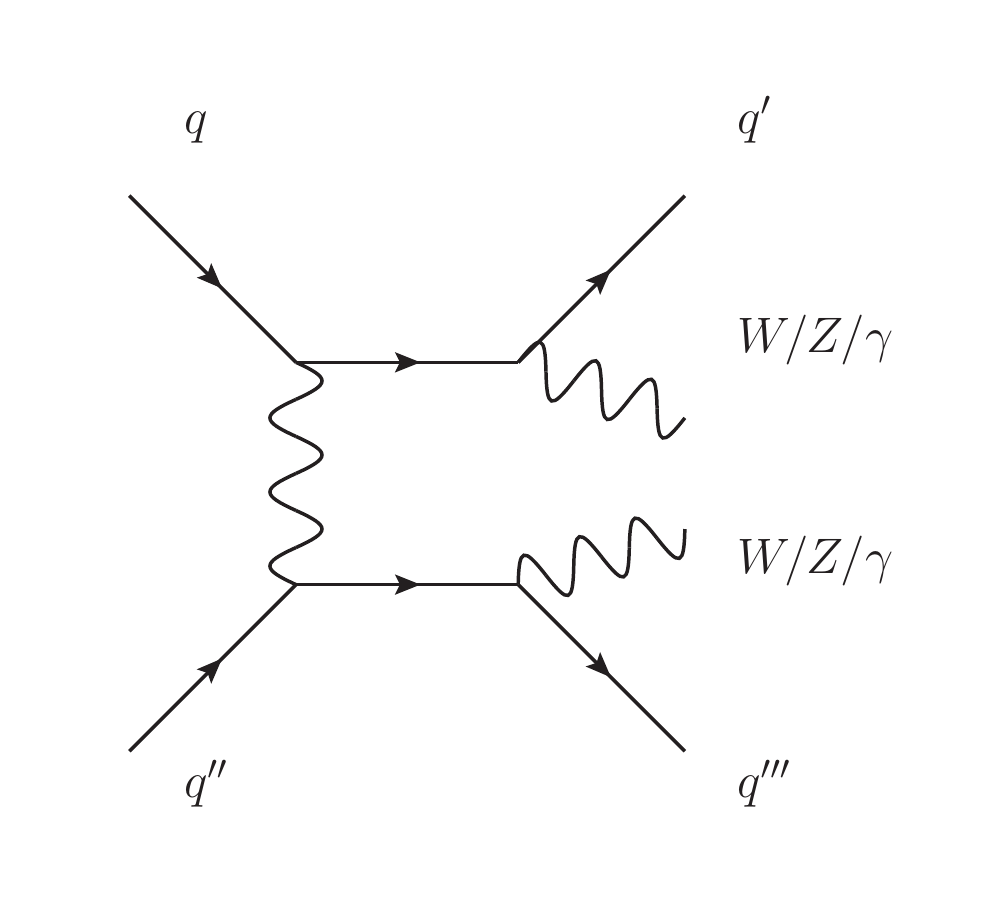}}\\
\subfloat[]{\includegraphics[width=0.35\columnwidth]{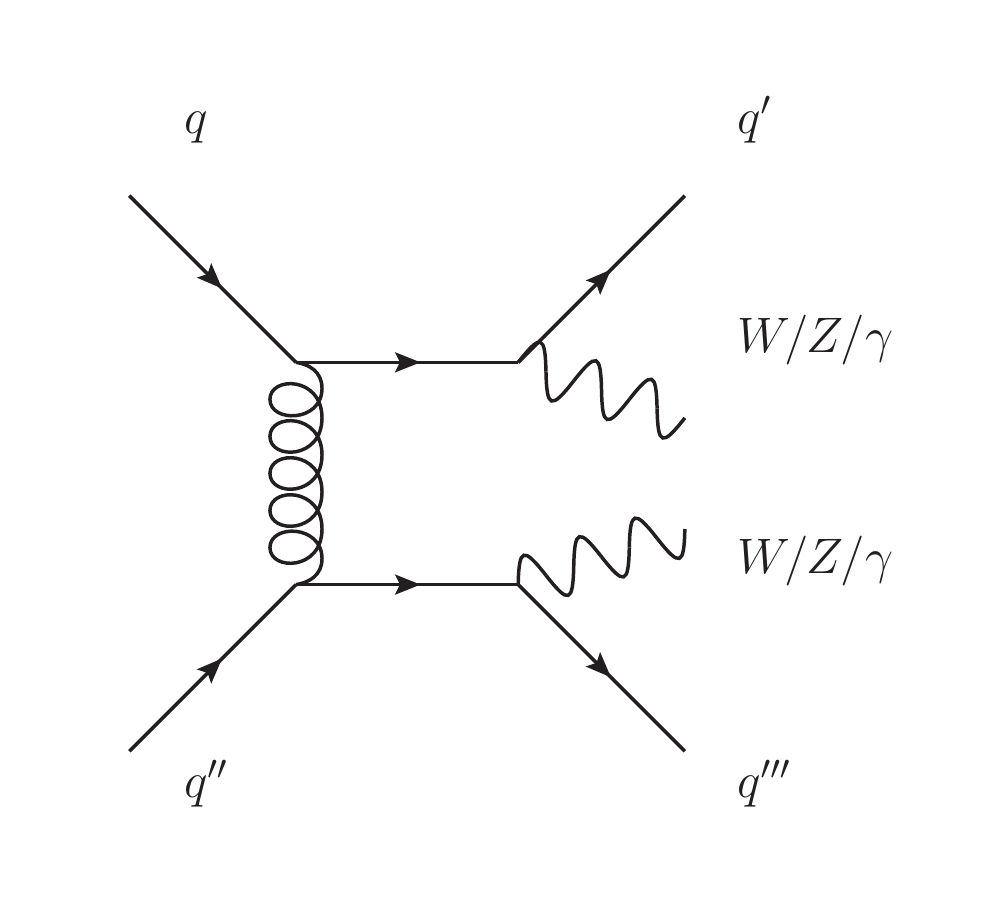}}
\subfloat[]{\includegraphics[width=0.35\columnwidth]{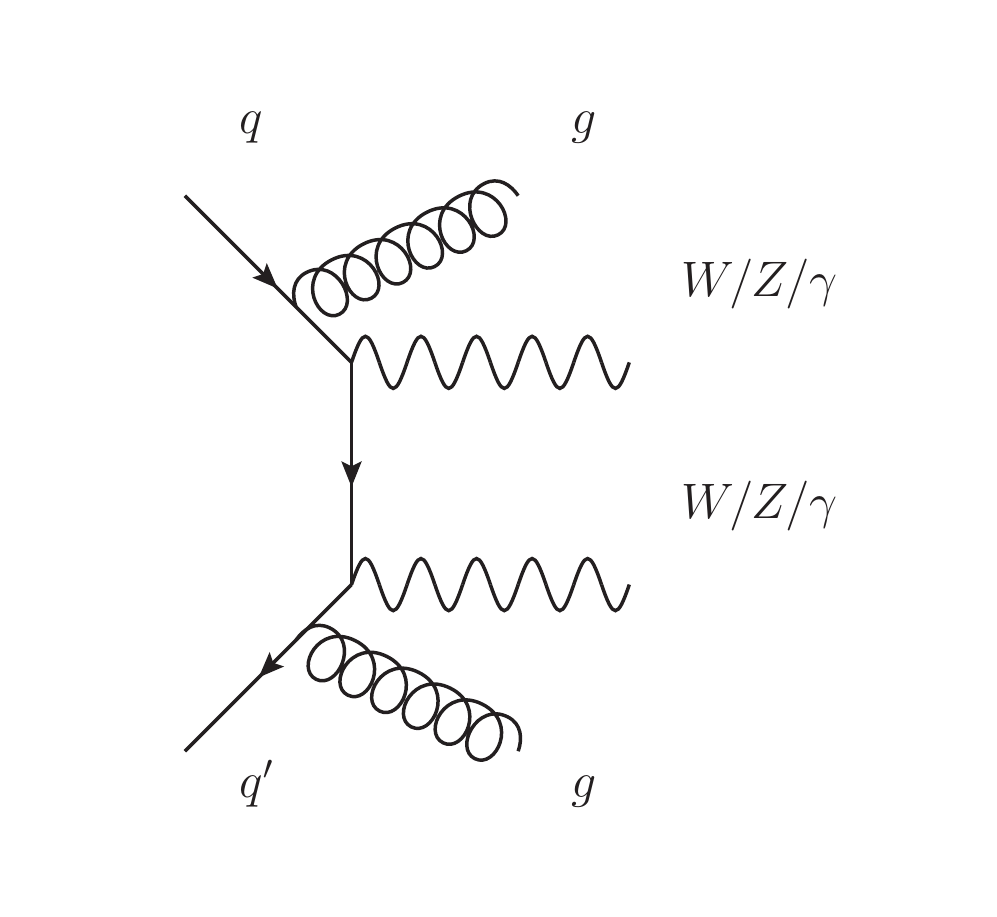}}
 
\caption{Representative Feynman diagrams of the processes relevant to this analysis: (a) quartic gauge coupling VBS signal, (b) triple gauge coupling, (c) electroweak non-VBS signal, QCD-induced backgrounds with (d) gluon exchange or (e) gluon radiation.}
\label{fig:FeynmanDiagrams}
\end{figure}

\section{ATLAS detector}
 
The ATLAS detector~\cite{PERF-2007-01, PIX-2018-001, ATLAS-TDR-2010-19}  is a multipurpose detector with a cylindrical geometry\footnote{ATLAS uses a right-handed coordinate system with its origin at the nominal interaction point (IP) in the centre of the detector and the $z$-axis along the beam direction. The $x$-axis points from the IP to the centre of the LHC ring, and the $y$-axis points upward. Cylindrical coordinates $(r,\phi)$ are used in the transverse $(x,y)$ plane, $\phi$ being the azimuthal angle around the beam direction. The pseudorapidity is defined in terms of the polar angle $\theta$ as $\eta = -\textrm{ln}[ \textrm{tan}(\theta/2)]$. } and nearly $4\pi$ coverage in solid
angle.
 
The collision point is surrounded by inner tracking detectors, collectively referred to as the inner detector (ID), located within a superconducting solenoid providing a $2$~T axial magnetic field, followed by a calorimeter system and a muon spectrometer (MS).
 
The inner detector provides precise measurements of charged-particle tracks in the pseudorapidity range $|\eta| < 2.5$.
It consists of three subdetectors arranged in a coaxial geometry around the beam axis: a silicon pixel detector, 
a silicon microstrip detector and a transition radiation tracker.
 
The electromagnetic (EM) calorimeter covers the region $|\eta|<3.2$ and is based on high-granularity, lead/liquid-argon (LAr) sampling technology.
The hadronic calorimeter uses a steel/scintillator-tile detector in the region $|\eta|<1.7$ and a copper/LAr detector in the region $1.5 < |\eta| < 3.2$.
The most forward region of the detector, $3.1 <|\eta| < 4.9$, is equipped with
a forward calorimeter, measuring electromagnetic and hadronic energies in copper/LAr and tungsten/LAr modules, respectively.

The muon spectrometer comprises separate trigger
and high-precision tracking chambers to measure the deflection of muons in a magnetic field generated by three
large superconducting toroidal magnets arranged with an eightfold azimuthal coil symmetry around the calorimeters.
The high-precision chambers cover the range $|\eta|< 2.7$.
The muon trigger system covers the range $|\eta|< 2.4$ with
resistive-plate chambers in the barrel and thin-gap chambers in the endcap regions.
 
A two-level trigger system~\cite{Aaboud:2016leb} is used to select events in real time. It consists of a hardware-based first-level (L1) trigger and a software-based high-level trigger. The latter employs algorithms similar to those used offline to identify electrons, muons, photons and jets.

\section{Signal and background simulation}

Signal and background processes are modelled using Monte Carlo (MC) simulations. All MC events were passed through the \textsc{Geant4}-based~\cite{Agostinelli:2002hh} ATLAS detector simulation~\cite{Aad:2010ah}. They are reconstructed and analysed using the same algorithm chain as the data. These events also include additional proton--proton interactions (pile-up) generated with \PYTHIA~$8.186$~\cite{Sjostrand:2007gs} using the \textsc{MSTW2008LO}~\cite{Martin:2009iq} parton distribution functions (PDFs)
and the A2 set of tuned parameters~\cite{ATL-PHYS-PUB-2012-003} (A2 tune).
The simulated events are reweighted to reproduce the mean number of interactions per bunch crossing observed during 2015 and 2016 data-taking. The average number of $pp$ interactions per bunch crossing is found to be around 13 in 2015 and 25 in 2016.
Scale factors are applied to simulated events to correct for the differences observed between data and MC simulation
in the trigger, reconstruction, identification, isolation and impact parameter efficiencies of photons, electrons and muons~\cite{Aaboud:2019ynx,PERF-2015-10}.
The photon energy, electron energy and muon momentum in simulated events are smeared to account for
differences in resolution between data and MC simulation~\cite{Aaboud:2018ugz,PERF-2015-10}.

In this analysis, the signal is the electroweak production of $Z(\rightarrow \ell \ell)\gamma jj$ (with $\ell$ = $e,\mu$),
with jets originating from the fragmentation of partons arising from electroweak vertices. Signal events ($\ell \ell\gamma jj$) were generated~\cite{ATL-PHYS-PUB-2017-005} at leading-order (\LO) accuracy (at order $\alpha_{\mathrm{EW}}^5$) using \textsc{MadGraph5\_aMC@NLO}~2.3.3~\cite{Alwall:2014hca} with no extra parton in the final state. The nominal renormalisation and factorisation scales were set to the transverse mass of the diboson system. The \textsc{NNPDF30} \LO PDF set~\cite{Ball:2014uwa} was used for the generation of the events, and the hadronisation and parton shower of the events was modelled using \PYTHIA~$8.212$ with the A14 tune~\cite{ATL-PHYS-PUB-2014-021}. An alternative model of hadronisation and parton shower is obtained by re-showering events with \Herwig~7.0.1~\cite{Bahr:2008pv,Bellm:2015jjp} using the \textsc{UEEE-5} tune that is based on the models discussed in Ref.~\cite{Gieseke:2012ft,Seymour:2013qka} in conjunction with the \textsc{CTEQ6L1} \LO PDF set~\cite{Pumplin:2002vw}.

An alternative sample of simulated events at \LO accuracy was generated using \Sherpa~2.2.4~\cite{Gleisberg:2008ta} with Comix~\cite{Gleisberg:2008fv} and merged with the \Sherpa parton shower~\cite{Schumann:2007mg}. The factorisation and renormalisation scales were set to the invariant mass of the diboson system. The \textsc{NNPDF30} next-to-next-to-leading order (NNLO) PDF set~\cite{Ball:2014uwa} was used for the generation of this sample.
 
The main background comes from the QCD-induced production of the $Z(\rightarrow \ell \ell)\gamma jj$ final state, which is estimated at leading order, i.e.\ at order $\alpha_\text{S}^{2}\alpha_{\mathrm{EW}}^3$.
The MC sample used to model this background was obtained with the \Sherpa~2.2.2 event generator for the $\ell \ell\gamma$ process with up to one parton at NLO generated with OpenLoops~\cite{Cascioli:2011va} and up to three partons at \LO using Comix. The different final-state multiplicities were merged with the \Sherpa parton shower using the \textsc{ME+PS@NLO} prescription~\cite{Hoeche:2012yf}. This sample was generated using the \textsc{NNPDF30} NNLO PDF set.
 
An alternative QCD sample was used to study the dependence of the kinematic distributions on the matrix-element generator. This sample was obtained with \textsc{MadGraph5\_aMC@NLO}~2.3.3 with up to one parton at \NLO for the $\ell \ell\gamma$ process. The second jet is obtained from the real emission at matrix-element level, i.e.\ with LO accuracy. The event generator was interfaced to \pythia~$8.212$ for the inclusion of the parton shower and hadronisation of the events; the FxFx merging prescription was used~\cite{Frederix:2012ps} with a merging scale of $20~\GeV$. The \textsc{NNPDF30} \NLO PDF set and the A14 tune were used for the generation.
 
Interference between the electroweak and QCD processes was estimated at \LO accuracy in QCD using the \textsc{MadGraph5\_aMC@NLO}~2.3.3 MC event generator with the \textsc{NNPDF30} \LO PDF set
including only contributions to the squared matrix-element at order $\alpha_\text{S}\alpha_{\mathrm{EW}}^4$. The event generator was interfaced to \pythia~$8.212$ for the parton showers and hadronisation of the events with the A14 tune.
These interference effects are found to be positive and are of the order of $3\%$ of the \zgjjew cross-section in the fiducial phase space studied by this analysis.

The second-largest background in this analysis is due to $Z+$jets processes. In this case, one of the jets is misidentified as a photon.
This contribution is estimated using a data-driven method, as explained in Section~\ref{sec:background}.
Cross-checks are performed using a \sherpa~2.2.1 MC sample~\cite{ATL-PHYS-PUB-2016-003}, produced with up to two jets at \NLO accuracy and up to four jets at \LO accuracy, using Comix and OpenLoops, and merged with the \Sherpa parton shower according to the \textsc{ME+PS@NLO} prescription. The \textsc{NNPDF30} NNLO PDF set was used for the generation of this sample. This sample is scaled to the NNLO predictions for inclusive $Z$ production using a normalisation factor~\cite{ATL-PHYS-PUB-2016-003} derived from \textsc{FEWZ}~\cite{Anastasiou:2003ds} and the \textsc{MSTW2008NNLO} PDF set~\cite{Martin:2009iq}.
 
The third-largest background affecting this analysis is from $t\bar{t}+\gamma$ events. These events were generated at \LO accuracy using the \textsc{MadGraph5\_aMC@NLO}~2.3.3 generator interfaced to \pythia{}~$8.212$ for the parton showering and hadronisation. The \textsc{NNPDF23} \LO PDF set and the A14 tune were used for the generation. The normalisation of the predictions is corrected for \NLO QCD effects~\cite{Melnikov:2011ta}.
 
All other backgrounds are smaller and are obtained from MC simulations. The most important of these are the $WZ$ background, which
was simulated with \sherpa~2.2.1 and the \textsc{NNPDF30} NNLO PDF set, and the $Wt$ background, which was simulated with the \powheg{}-Box~\cite{powheg,powheg1:2004,Alioli:2010xd} generator interfaced to \pythia{}~$6.428$~\cite{Sjostrand:2006za} and the \textsc{perugia}~2012 tune~\cite{Skands:2010ak}. The \textsc{CT10} \NLO PDF set~\cite{Lai:2010vv} was used for the generation of this sample.

\section{Event selection}
 
\label{sec:selection}
Events are selected if they were recorded during stable beam conditions and if they satisfy detector and data quality requirements, which include all relevant subdetectors functioning normally. Events were collected using single-lepton or dilepton triggers~\cite{Aaboud:2016leb,Aad:2019wsl} that require, respectively, at least one or two electrons or muons.
 
For single-lepton triggers, the transverse momentum ($\pt$) threshold in 2015 was $24~\GeV$ for electrons and $20~\GeV$ for muons satisfying a loose isolation requirement based only on ID track information. In 2016, due to the higher instantaneous luminosity, these thresholds were increased to $26~\GeV$ for both the electrons and muons, and tighter isolation requirements were applied.
Inefficiencies for leptons with large transverse momenta were reduced by including additional electron and muon
triggers that do not include any isolation requirements; these had transverse momentum thresholds of $60~\GeV$ and $50~\GeV$, respectively.
Finally, a single-electron trigger requiring $\pT>120~\GeV$ with less restrictive electron identification criteria was used to increase the selection efficiency for high-$\pt$ electrons. The threshold was increased to $140~\GeV$ in $2016$.
In 2015, the dilepton trigger $\pT$ threshold was $12~\GeV$ for electrons, and either $10~\GeV$ for muons in conjunction with two muons producing L1 triggers, or $18~\GeV$ and $8~\GeV$ in the case that only one of the two muons produced an L1 trigger. In 2016, these thresholds were raised to $17~\GeV$ for electrons, $14~\GeV$ for the symmetric dimuon trigger, and $22~\GeV$ and $8~\GeV$ for the asymmetric muon trigger.
The combined efficiency of these triggers is close to $100\%$ for the events that pass the kinematic requirements described below. To ensure that the trigger efficiency is well determined, the lepton candidates must be matched to the leptons that are selected by the trigger and have a transverse momentum at least $1~\GeV$ above the online threshold.
 
Events must have a primary vertex with at least two charged-particle tracks which must be compatible with the \pp\ interaction region. In cases where multiple vertices are reconstructed in a single event, the vertex with the highest sum of the $\pT^{2}$ of the associated tracks is selected as the production vertex of the $Z\gamma$ system.

Muons are reconstructed by matching tracks in the muon spectrometer to tracks in the inner detector. Candidate muons are required to satisfy the `medium' set of identification criteria~\cite{PERF-2015-10} based on the number of detector hits and $\pt$ measurements in the ID and the MS. The efficiency of selecting such muons averaged over \pt and $\eta$ is larger than $98\%$.
Both the ID and MS measurements are used to compute the muon momentum. The measurement also takes into account the energy loss in the calorimeters. Muons are used in the analysis if they satisfy $\pt>20~\GeV$ and $|\eta|<2.5$.
 
Electrons are reconstructed by matching energy clusters in the electromagnetic calorimeter to an ID track. Candidate electrons are required to pass a `medium' likelihood requirement that is built from information about the electromagnetic shower shape in the calorimeter, track properties and the track-to-cluster matching~\cite{Aaboud:2019ynx}. The efficiency of selecting such electrons is of about $90\%$.
Calorimeter energy and the track's direction are combined to compute the electron momentum. Electrons are required to satisfy $\pt>20~\GeV$ and $|\eta|<2.47$.
 
To ensure that electron and muon candidates originate from the primary vertex, the significance of the track's transverse impact parameter relative to the beam line must satisfy $|{d_{0}/\sigma_{d_{0}}}| <3\ (5)$ for muons (electrons), and the longitudinal impact parameter $z_{0}$, the difference between the value of $z$ at the point on the track at which $d_{0}$ is defined and the longitudinal position of the primary vertex, is required to satisfy $|z_0\cdot \sin(\theta)|<0.5$~mm.
 
Electrons and muons are required to be isolated from hadronic activity by imposing maximum energy requirements in cones in the $\eta$--$\phi$ plane, defined using the $\Delta R= \sqrt{(\Delta{}\eta)^{2}+(\Delta{}\phi)^{2}}$ distance, around their direction of flight, using calorimeter-cluster and ID-track information, excluding the electron or muon momentum. They are selected following the `gradient loose' requirements~\cite{PERF-2015-10,Aaboud:2019ynx}. The muon isolation criterion imposes a fixed upper limit in a cone of size $\Delta R = 0.2$ for calorimeter-based isolation and in a cone of size $\Delta R = 0.3$ for ID-based isolation, with a selection efficiency above $90\%$ for $\pT>20~\GeV$ and at least $99\%$ for $\pT>60~\GeV$. For electrons, the requirements vary with \pt{}, allowing a selection efficiency of at least $90\%$ for $\pT>25~\GeV$ and at least $99\%$ for $\pT>60$~\GeV. For electrons, both isolation variables are evaluated with a cone of size $\Delta R = 0.2$.
 
Photon candidates are reconstructed from clusters of energy deposited in the electromagnetic calorimeter and are classified as unconverted (clusters without a matching track or matching reconstructed conversion vertex in the ID) or converted (clusters with a matching reconstructed conversion vertex or a matching track consistent with originating from a photon conversion).
Photons are identified using a cut-based selection relying on shower shapes measured with the electromagnetic calorimeter. Nine discriminating variables are used,
built from the distribution of energy in different layers of the EM calorimeter and information about energy leaking into the hadronic calorimeter. The shower shape values in the simulation are corrected to improve their agreement with the shower shapes in data.
Photons must satisfy a `tight' requirement~\cite{Aaboud:2018yqu}.
Photons selected in this analysis must satisfy $E_\mathrm{T}>15~\GeV$, where $E_\mathrm{T}$ is the photon transverse energy, and the pseudorapidity of the cluster must be in the range $ \abseta < 1.37$ or $1.52 < \abseta < 2.37$, to avoid the transition region between barrel and endcap calorimeters.
Photon candidates are required to be isolated according to calorimeter-cluster and ID-track information in a cone of size $\Delta R = 0.2$, excluding the photon energy. The requirement varies with $E_\mathrm{T}$ following the `fixed loose' requirement~\cite{Aaboud:2018yqu}, providing an average efficiency of about $95\%$. The calorimetric photon isolation is corrected to account for leakage and the contribution from the underlying event and pile-up transverse energy~\cite{Aaboud:2019ynx}.

Jets are reconstructed from clusters of energy depositions in the calorimeter~\cite{PERF-2014-07} using the anti-$k_{t}$ algorithm~\cite{antikt,Fastjet} with a radius parameter $R=0.4$. Calibration of the jet energy is based on simulation and {\textit{in situ}} methods from data~\cite{Aaboud:2017jcu}. Jets originating from non-collision backgrounds or detector noise are removed~\cite{ATLAS-CONF-2015-029}. Pile-up jets are suppressed using a multivariate combination of track-based variables in the ID acceptance ($\abseta<2.5$)~\cite{Aad:2015ina}.
All jets considered must have $\pt>25~\GeV$ and must be reconstructed in the pseudorapidity range $|\eta|<4.5$.
 
Jets containing a $b$-hadron are identified in the ID volume using a multivariate algorithm~\cite{PERF-2012-04,ATL-PHYS-PUB-2016-012} that uses the impact parameters and secondary vertices of the tracks contained in the jet. The selection is done using a working point chosen to provide a $70\%$ selection efficiency for $b$-jets in an inclusive $t\bar{t}$ MC sample. The 70\% working point has rejection factors of 10 and 400 for charm and light-flavour jets, respectively. Correction factors are applied to the simulated event samples to compensate for differences between data and simulation in the flavour-tagging efficiencies for $b$-jets, $c$-jets and light-flavour jets. The correction for $b$-jets is derived from $t\bar{t}$ events with final states containing two leptons, and the corrections are consistent with unity with uncertainties at the level of a few percent over most of the jet $\pt$ range.
 
To avoid cases in which a lepton, photon, or jet is reconstructed as two separate final-state objects, several steps are followed to remove such overlaps. Bremsstrahlung radiation by a muon can result in ID tracks and a calorimeter energy deposit that are reconstructed as an electron candidate. Therefore, in cases for which an electron candidate and a muon candidate share an ID track, the object is considered to be a muon, and the electron candidate is rejected. The overlap of objects is measured using the $\Delta R$ distance. Due to the isolation requirements placed on electron candidates, any jets that closely overlap an electron candidate within a conical region $\Delta R <0.2$ are likely to be reconstructions of the electron and so are rejected. When jets and electrons are found within a larger hollow conical region $0.2<\Delta R <0.4$, it is more likely that a real hadronic jet is present and that the electron is a non-prompt constituent of the jet, arising from the decay of a heavy-flavour hadron. Hence an electron candidate found within $\Delta R <0.4$ of any remaining jet is rejected. Muons can be accompanied by a hard photon due to bremsstrahlung or collinear final-state radiation, and the muon--photon system can then be reconstructed as both a jet and a muon candidate. Non-prompt muons can arise from decays of hadrons in jets; however, these muons are associated with a higher ID track multiplicity than those accompanied by hard photons. In order to resolve these ambiguities between nearby jet and muon candidates, first any jets having fewer than three ID tracks and within $\Delta R <0.4$ of any muon candidate are rejected, then any muon candidates within $\Delta R <0.4$ of any remaining jet are rejected. The ambiguity between photons and leptons is resolved by rejecting any photons found to be within $\Delta R <0.4$ of an electron or a muon. Jets found within $\Delta R <0.4$ of any photons are rejected.

Events are required to contain exactly two lepton candidates (electrons or muons) with the same flavour and opposite charge and at least one photon candidate satisfying the selection criteria described above. When there are two or more photons selected, the one with the highest $E_\mathrm{T}$ is used for further processing.
 
The invariant mass of the two leptons, $m_{\ell \ell}$, must be at least $40~\GeV$. The sum of the dilepton mass and the three-body $\ell \ell\gamma$ invariant mass ($m_{\ell \ell}+m_{\ell \ell\gamma}$) is required to be larger than $182~\GeV$, which is twice the $Z$ boson mass. This requirement ensures that the three-body invariant mass is larger than the $Z$ boson mass, thus suppressing cases where the $Z$ boson decays into ${\ell \ell}\gamma$~\cite{Aaboud:2017pds,Aad:2019gpq}.

To select electroweak processes, events are further required to include at least two `VBS-tagging' jets. A symmetric selection of $\pt>50~\GeV$ is used to select both tagging jets, which are classified as leading and subleading in transverse momentum. These two jets are required to have an invariant mass $m_{jj}>150~\GeV$, in order to minimise the contamination from triboson background in which a hadronically decaying $W$ or $Z$ boson is produced in association with a $Z$ boson and a photon, i.e.\ $Z\gamma + Z/W(\rightarrow j j)$.
 
The final VBS signal region (SR) is defined by requiring that the pseudorapidity difference between the two leading jets $|\Delta \eta (jj)|$ be larger than 1;  that no $b$-tagged jet be present in the event; and that the centrality of the $\ell\ell\gamma$ system relative to the tagging jets, defined as
\begin{equation}
\zeta(\ell\ell\gamma) = \left| \frac{y_{\ell\ell\gamma} - (y_{j_{1}} + y_{j_{2}})/2}{(y_{j_{1}} - y_{j_{2}})} \right|,
\label{eqn:centrality}\\
\end{equation}
be lower than 5, where $y$ corresponds to the rapidity, and $j_{1}$ and $j_{2}$ label the selected leading and subleading jets.

\section{Background estimation}
 
\label{sec:background}
Irreducible backgrounds are modelled kinematically by MC simulations, and their normalisation is evaluated from the data. Reducible backgrounds, in which a photon originates from a hadron decay, are determined using data-driven techniques. Events in which one or more leptons originate from a hadronic jet are suppressed by the the stringent lepton selection. They are found to be negligible and are not considered here.
 
The main reducible background originates from $Z+$jets events, in which one hadronic jet is misidentified as a photon.
This background is not well modelled by the MC simulation and, therefore, is estimated with data using a two-dimensional sideband method~\cite{Aad:2010sp} similar to that applied in a previous analysis with Run~1 data~\cite{Aaboud:2017pds}.
The shapes of the distributions of $Z+$jets events are obtained using data, after applying all the signal requirements, but reversing the photon identification and isolation criteria~\cite{Aaboud:2018yqu}. The normalisation of this background is estimated in a control region designed to maximise the number of events available in the data while being close to the signal region; all signal region requirements are applied, except the transverse momentum requirement for the jets, which is lowered to $30~\GeV$, and the dijet invariant mass requirement, which is less than $150~\GeV$. The ratio of $Z+$jets to \zgjjqcd events is computed in this region, since it is found to be the same in this control region and in the signal region, in both data and MC simulation, and used to extrapolate the normalisation of $Z+$jets events to the signal region. To estimate both the shape and the normalisation of this background, the contamination from events containing real photons is estimated and corrected using MC predictions.
 
The most important background is irreducible and comes from the QCD-induced production of \zg{} events in association with two jets. The normalisation of this background is constrained by the data, as explained in Section~\ref{sec:analysis}.
 
The other main irreducible background arises from $t\bar{t} + \gamma$ events. This process is also constrained by the data in a dedicated control region which is referred to as $b$-Control Region ($b$-CR). This control region is obtained by imposing all SR selections but requiring the presence of at least one reconstructed $b$-tagged jet in the event.
 
The contribution of backgrounds due to the production of $Z$+jets and $\gamma$+jets events in two different overlaid interactions is evaluated using a data-driven estimate following the method described in Ref.~\cite{Aad:2019gpq} and found to be negligible.
 
Backgrounds from other sources are reducible and are evaluated using MC simulations. The most important contribution comes from $WZ$ events, in which a charged lepton is misidentified as a photon, and from $Wt$ events.

\section{Signal extraction procedure}
 
\label{sec:analysis}
 
A boosted decision tree (BDT), as implemented in the TMVA package~\cite{Hocker:2007ht}, is used to separate the electroweak signal from all the backgrounds. The BDT is trained and the set of variables retained as input to the BDT is optimised on simulated events to separate $\zgjjew$ events from all background processes,  excluding $Z+$jets, and to maximise the signal-to-background ratio. In total, 13 kinematic variables are combined into one discriminant that can take any value in the range $[-1,1]$. The final binning of the BDT discriminant, or BDT score, is optimised to improve the sensitivity of the analysis.
 
Some of the variables used in the BDT training are related to the kinematic properties of the two tagging jets: the invariant mass of these two jets, $m_{jj}$, the difference in pseudorapidity between these two jets, $\Delta \eta_{jj}$, and the pseudorapidity and $\pT$ of the leading jet. Other variables are related to the kinematic properties of the photon and the $Z$ boson: the transverse momenta of the leading lepton, the $\ell\ell$ system and the $\ell\ell\gamma$ system, and the $\ell\ell$ and $\ell\ell\gamma$ invariant masses. Another set of variables is related to the correlation of the two tagging jets and the $Z\gamma$ system: the distance in the $\eta$--$\phi$ plane between the $Z\gamma$ system and the two-jet system, $\Delta R(Z\gamma, jj)$;
the smallest distance in the $\eta$--$\phi$ plane between a photon and a jet, among all possible combinations, Min($\Delta R (\gamma, j)$); the difference in azimuth between the $Z\gamma$ system and the two-jet system, $\Delta \phi (Z\gamma, jj)$ and the centrality $\zeta(\ell\ell\gamma)$ as defined in Eq.~(\ref{eqn:centrality}).

The modelling by MC simulations of the distribution shapes and the correlations between all input variables for the BDT is verified in the signal region. Compatibility within the uncertainties is observed for all distributions, as exemplified by Figures~\ref{fig:BDTInputs}(a) and~\ref{fig:BDTInputs}(b), except for the high-mass tail of $m_{jj}$ as shown in Figure~\ref{fig:BDTInputs}(c), as was also observed in previous analyses of electroweak processes~\cite{Aad:2014dta,Aaboud:2017fye,Aaboud:2017emo}. This mismodelling does not impact the signal significance nor the uncertainty in the measurement reported in Section~\ref{sec:result}.

\begin{figure}[!htbp]
\begin{center}
\subfloat[]{\includegraphics[width=0.45\textwidth]{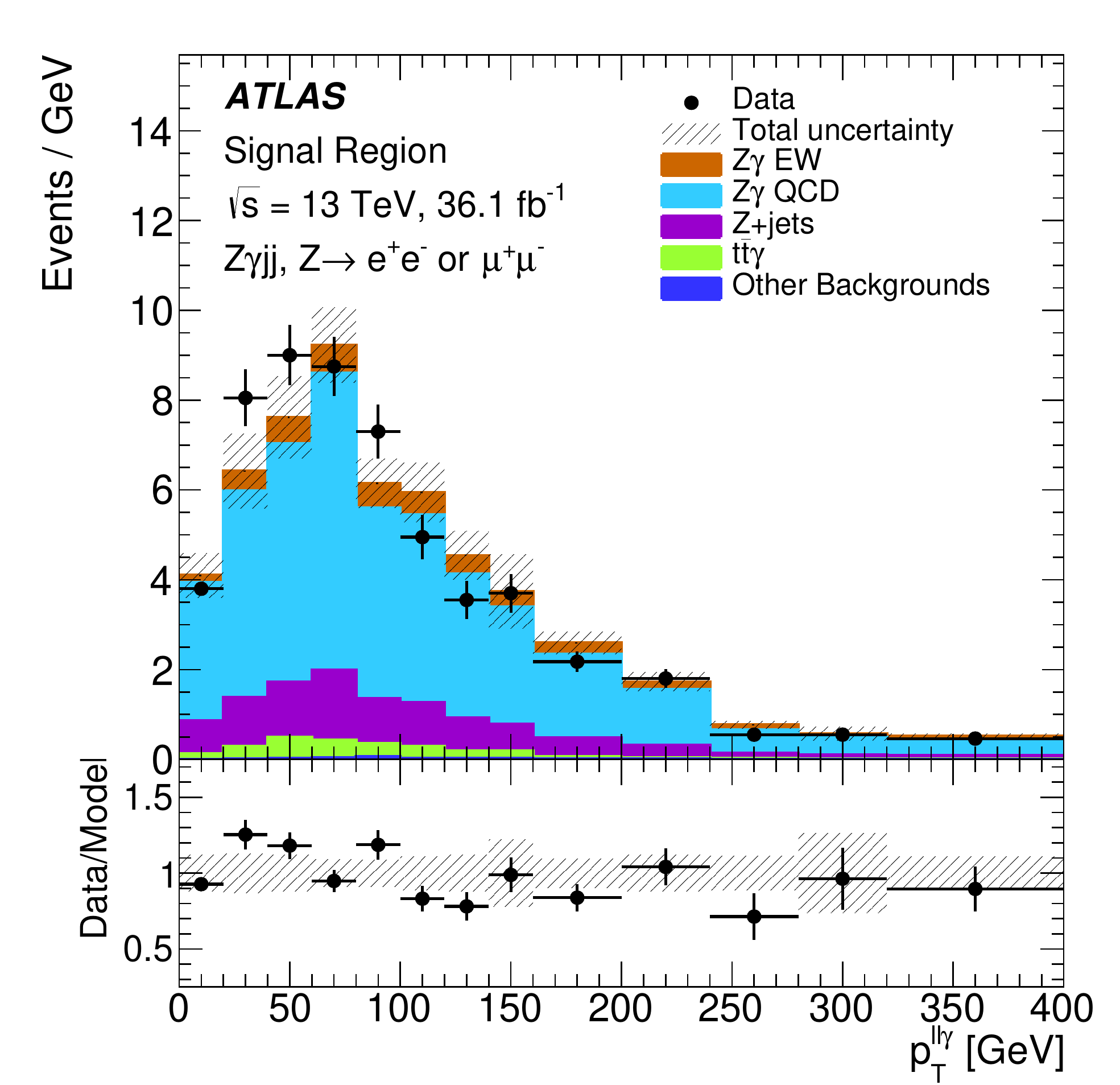}}
\subfloat[]{\includegraphics[width=0.45\textwidth]{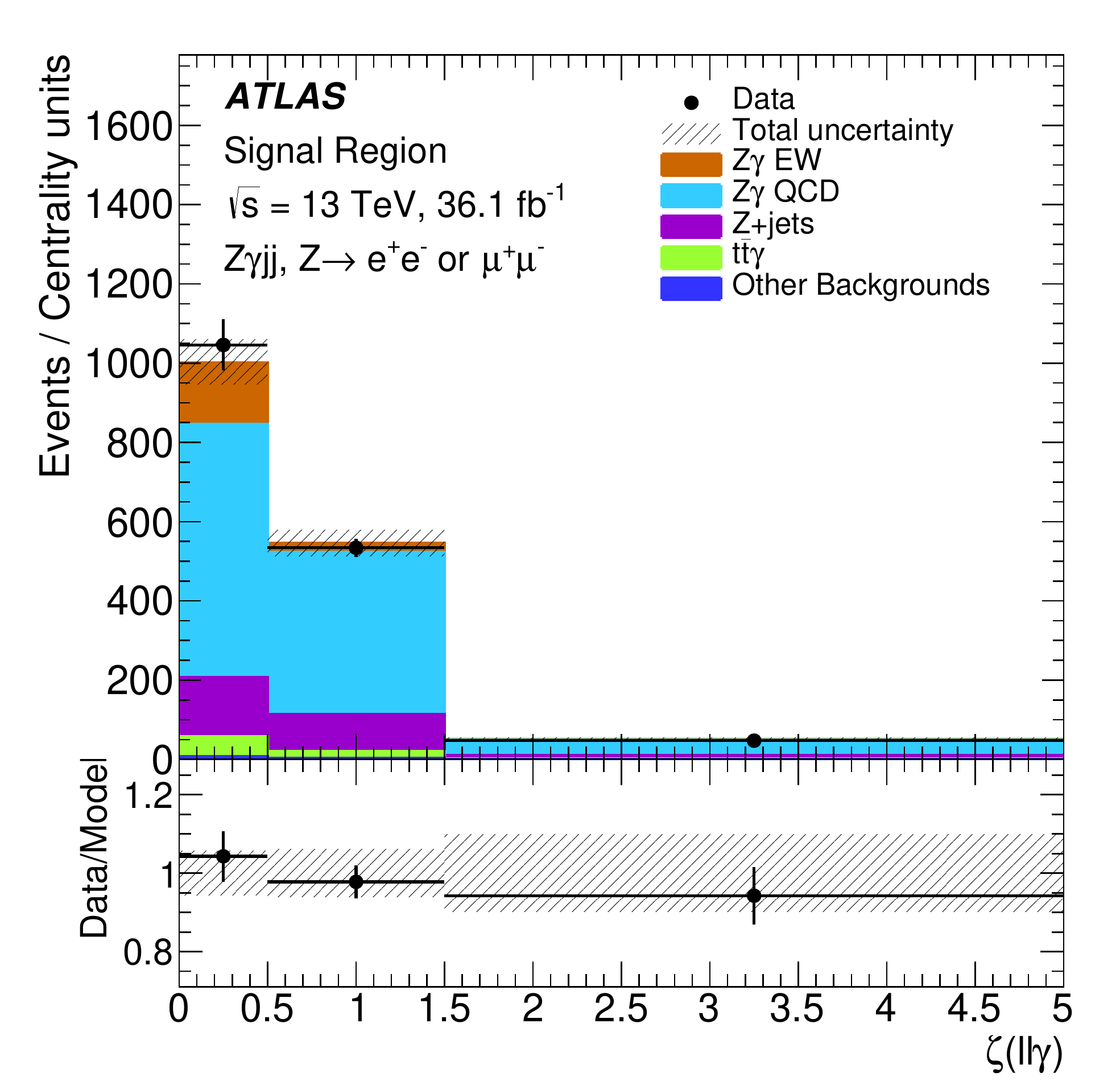}}\\
\subfloat[]{\includegraphics[width=0.45\textwidth]{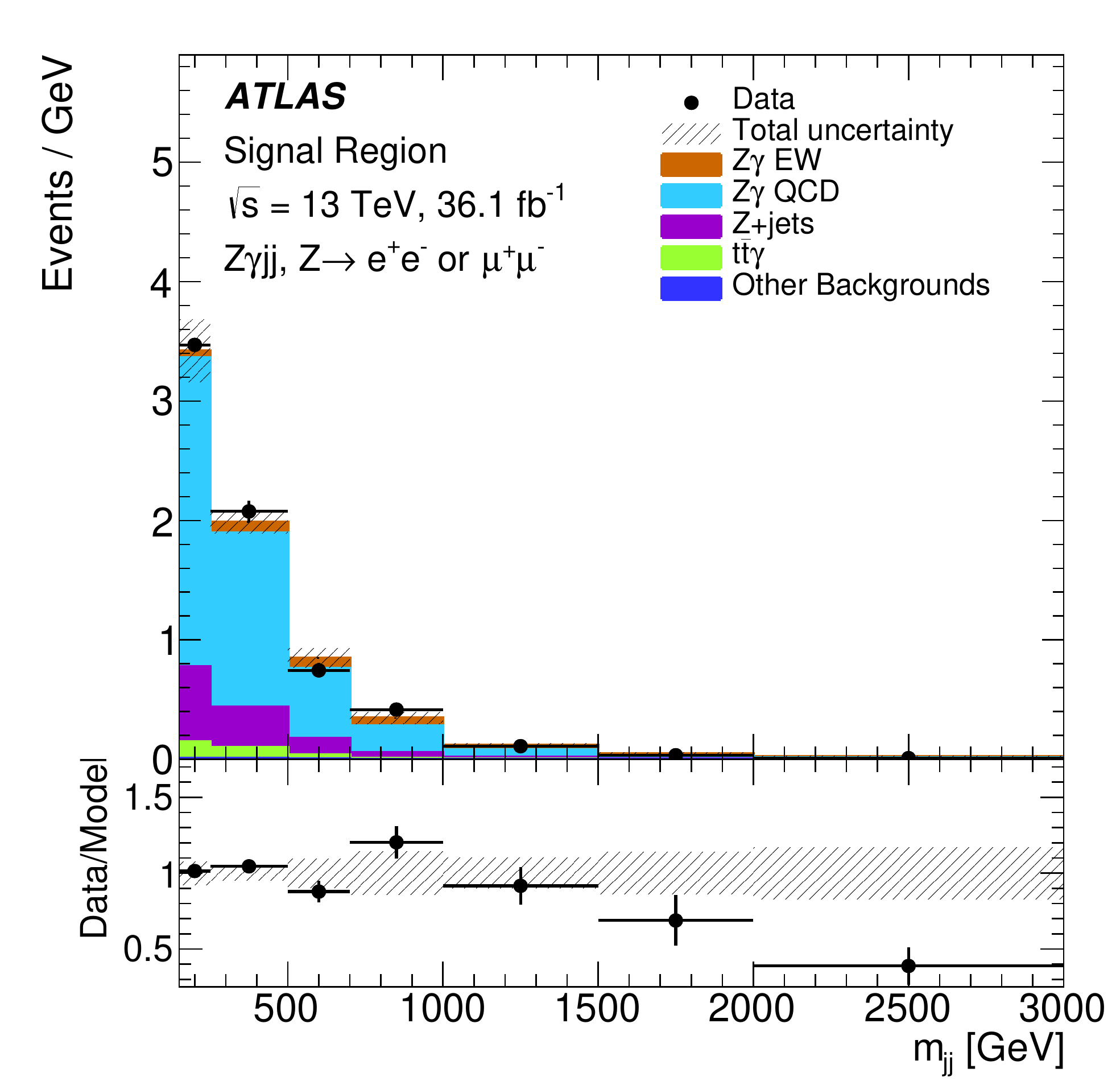}}
\caption{Post-fit distributions of (a) the transverse momentum of the $\ell\ell\gamma$ system, (b) the centrality of the $\ell\ell\gamma$ system relative to the tagging jets, and (c) $m_{jj}$ in the signal region.
The uncertainty band around the expectation includes all systematic uncertainties and takes into account their correlations as obtained from the fit. Events beyond the upper limit of the histogram are included in the last bin of the distribution.}
\label{fig:BDTInputs}
\end{center}
\end{figure}

The BDT score distribution in the signal region is used to measure the \zgjjew signal fiducial cross-section with a maximum-likelihood fit, and to determine the significance of the signal. In this fit the $ee\gamma{}jj$ and $\mu\mu\gamma{}jj$ final states are combined. An extended likelihood function is built from the product of two likelihoods corresponding to the BDT score distribution in the \zgjj SR, and the multiplicity of reconstructed $b$-jets in the $b$-CR. The yields of the $t\bar{t}+\gamma$ simulated events are constrained by data using this last distribution. The normalisation of the \zgjjqcd sample is also constrained by the data, directly in the signal region, since this background dominates the distribution at low BDT score values. The normalisations of these backgrounds are introduced in the likelihood fit as unconstrained nuisance parameters, labelled as $\mu_{\zgjjqcd}$ and $\mu_{t\bar{t}+\gamma}$ for the
\zgjjqcd and $t\bar{t}+\gamma $ backgrounds, respectively. These parameters are relative normalisation factors, defined with respect to the SM predictions.
For these backgrounds, the kinematic distributions are obtained from MC simulations.
Both the shape and the normalisation distributions of $Z+$jets background are estimated from data, as described in Section~\ref{sec:background}.
The other irreducible backgrounds are determined from MC simulations.
Background normalisations and shapes can vary within the uncertainties, constrained by Gaussian distributions as described in Section~\ref{sec:systematics}.
 
The fiducial cross-section is derived using the signal strength parameter $\mu_{\zgjjew}$:
 
\[
\mu_{\zgjjew} = \frac{ N_{\zgjjew}^{\mathrm{data}} } {N_{\zgjjew}^{\mathrm{MC}} }\; ,
\]

where $N_{\zgjjew}^{\mathrm{data}}$ is the number of signal events measured in the data, while $N_{\zgjjew}^{\mathrm{MC}}$ is the number of signal events predicted by the \textsc{MadGraph5\_aMC@NLO}~2.3.3 MC simulation.
 
The observed cross-section ($\sigma_{\zgjjew}^{\mathrm{fid.\,, data}}$) is then obtained by multiplying the signal strength $\mu_{\zgjjew}$ by the \textsc{MadGraph5\_aMC@NLO}~2.3.3 MC cross-section prediction in the fiducial region ($\sigma_{\zgjjew}^{\mathrm{fid.,\, MC}}$ ).
Because the effect of interference between the \zgjjqcd and the \zgjjew processes is not accounted for in the \zgjjqcd contribution, the observed cross-section $\sigma_{\zgjjew}^{\mathrm{fid.\,}}$ formally corresponds to electroweak production plus the interference effects.

\section{Systematic uncertainties}
 
\label{sec:systematics}
Several sources of systematic uncertainty in the signal and background processes can affect the cross-section measurement.
For the background and signal processes, all systematic uncertainties that affect the shapes of the BDT score and $N_{b\textrm{-jets}}$ distributions are considered.
In addition, all systematic uncertainties that affect the acceptance and normalisation of the distributions are considered for all processes, except for the theory uncertainties in the signal process, where only the effect on the acceptance is considered.
Systematic uncertainties in the shapes of distributions are not applied if they are consistent with statistical fluctuations.
 
The uncertainties in the theoretical modelling of the signal and backgrounds that are estimated from MC simulations can affect the cross-section measurement. The uncertainties due to the knowledge of the PDF and the $\alpha_\text{S}$ value used in the generation of the MC samples are determined using the \textsc{PDF4LHC} prescription~\cite{Butterworth:2015oua}. They are found to be of the order of 2\% for the normalisation of the QCD and $t\bar{t}\gamma$ samples. The uncertainties due to missing higher-order QCD corrections are evaluated using an envelope of the largest deviations obtained by varying the renormalisation and factorisation scales independently by factors of two and one-half. These uncertainties are large for the \zgjjqcd and $t\bar{t}+\gamma$ backgrounds and their impact on the normalisation is $^{+30}_{-20}$\%. The shape uncertainties due to the missing higher-order QCD corrections are of 2--4\% for the signal and the \zgjjqcd and $t\bar{t}+\gamma$ backgrounds.
 
For the signal and $t\bar{t}+\gamma$ background, uncertainties due to the parton shower and underlying-event modelling are evaluated using dedicated tune variations~\cite{ATL-PHYS-PUB-2014-021} and by re-showering events with \Herwig~7.0.1 instead of \PYTHIA~$8.212$. These uncertainties affect the shape of the distribution by at most $5\%$ at large values of the BDT score and are negligible at low values of the BDT score. For the signal this uncertainty takes into account a problem with color flow connection in VBS-like topology that was observed and reported in Ref.~\cite{Ballestrero:2018anz,ATL-PHYS-PUB-2019-004} in the parton shower models of \PYTHIA~$8.212$ and \SHERPA~2.2.4. For the \zgjjqcd background, the modelling uncertainties due to the parton showers, underlying-event and matrix elements are estimated by comparing the predictions of the \sherpa~2.2.2 and \textsc{MadGraph5\_aMC@NLO}~2.3.3 MC generators. The difference between these two predictions is taken as a systematic uncertainty applied as both a positive and a negative variation. The effects in the BDT score distribution and the $N_{b\text{-jets}}$ distribution due to these two set-ups are taken as uncertainties. The main impact is on the BDT score distribution with an effect ranging from $-5\%$ to $20\%$ at low and high values of the BDT score, respectively.
 
As discussed in Section~\ref{sec:analysis}, the interference between the electroweak signal and the \zgjjqcd process is not included in the fit. However, this interference can distort the shape of the \zgjjew template. This effect is estimated using the \textsc{MadGraph5\_aMC@NLO}~2.3.3 MC generator at LO in QCD, and is taken as a systematic uncertainty in the shape of the signal. The size of this effect ranges from $5\%$ to $2\%$ at low and high values of the BDT score, respectively.

Another source of uncertainty arises from the sizes of the MC samples and data sample in the regions used in the analysis to model the BDT score and $N_{b\textrm{-jets}}$ distributions. The statistical uncertainty in the shape of the BDT score for \zgjjqcd events ranges from $5\%$ to $13\%$ at low and high values of the BDT score, respectively, and is about $2\%$ for \zgjjew events. For the $Z+$jets background, the statistical uncertainty dominates the shape uncertainty due to the size of the data sample; it is about $10\%$ at low BDT score values, and up to $50\%$ at high values, while uncertainties from other sources are negligible.  The statistical uncertainty in the shape of $N_{b\text{-jets}}$ is about $25\%$ for \zgjjqcd and is about $3\%$ for $t\bar{t}\gamma$ events.
 
Other systematic uncertainties originate from the reconstruction, identification, and energy calibration of electrons, photons, muons, and jets.
The largest of these uncertainties is due to the jet energy scale calibration. The uncertainties due to the jet energy resolution and to the suppression of pile-up jets are also considered, as described in Ref.~\cite{PERF-2011-04}. The total effect of the uncertainties related to jet reconstruction and calibration on the normalisation of the \zgjjqcd background is about $8\%$, and it is about $4\%$ for \zgjjew events in the signal region. The impact on the shape is largest for the \zgjjqcd events and ranges from $2\%$ to $18\%$ at low and high BDT values, respectively.
The uncertainty due to the heavy-flavour tagging efficiency amounts to $9\%$ ($2\%$) in normalisation in the signal ($b$-CR) region for the $t\bar{t}+\gamma$ background, whereas the impact on other backgrounds and on the shape of the template distributions is found to be negligible.
Uncertainties in the lepton identification, reconstruction, isolation requirements, trigger efficiencies, energy scale and energy resolution are determined by using $Z \to \ell \ell$ events~\cite{Aaboud:2019ynx,PERF-2015-10,PERF-2013-05}.
The largest experimental uncertainty associated with the photons is the photon identification efficiency~\cite{Aaboud:2018yqu}.
The photon and lepton uncertainties affect only the normalisation of \zgjjew, \zgjjqcd, and $t\bar{t}+\gamma$ processes.
In total, these effects are of $3\%$ to $4\%$ in the signal region.

A $20\%$ yield uncertainty is assigned to the $Z+$jets reducible background estimate. This uncertainty accounts for the number of events in the control regions used in the two-dimensional sideband measurement, and for the correlation between photon identification and isolation requirements.
A $20\%$ yield uncertainty is assigned to the other backgrounds estimated from simulations.
 
A variation in the pileup reweighting of MC samples is included to cover the uncertainty on the ratio between the predicted and measured inelastic cross-section~\cite{STDM-2015-05}. The resulting uncertainty in the measured fiducial cross-section is 5\%.

The uncertainty in the combined 2015--2016 integrated luminosity is $2.1\%$~\cite{ATLAS-CONF-2019-021}, obtained using the LUCID-2 detector \cite{LUCID2} for the primary luminosity measurements.

The effect of each of these uncertainties on the fiducial cross-section measurement is shown in Table~\ref{table:sysOnMu_summary}. The individual sources are grouped into either theoretical or experimental categories. The largest uncertainties are due to the jet reconstruction and calibration, followed by the uncertainty arising from the sizes of the MC samples, and the theoretical uncertainties in the modelling of the \zgjjew signal and of the \zgjjqcd background. Uncertainties affecting only the normalisation of the \zgjjqcd and $t\bar{t}\gamma$ background have almost no impact on the \zgjjew cross-section measurement since the corresponding normalisation parameters are constrained by the fit to the data as explained in Section~\ref{sec:analysis}.
 
\begin{table}[!htbp]
\renewcommand{\arraystretch}{1.1}
\caption{Relative uncertainties in the measured fiducial cross-section $\sigma^{\mathrm{fid.}}_{\zgjjew}$. The uncertainties are expressed in percentages. The correlations between these uncertainties are taken into account in the computation of the total uncertainty.}
\label{table:sysOnMu_summary}
\centering
\begin{tabular}{lc}
\toprule
Source & Uncertainty [\%]\\
\midrule
Statistical                              			& $^{+19}_{-18}$\\
\midrule
\zgjjew theory modelling                 			&  $^{+10}_{-6}$\\
\zgjjqcd theory modelling                			&  $\pm 6$ \\
$t\bar{t} +\gamma$ theory modelling      			&  $\pm 2$ \\
\zgjjew and \zgjjqcd interference        			&  $^{+3}_{-2}$ \\
\midrule
Jets                                     			&  $\pm 8$ \\
Pile-up                                  			&  $\pm 5$ \\
Electrons                                			&  $\pm 1$ \\
Muons                                    			&  $^{+3}_{-2}$\\
Photons                                  			&  $\pm 1$ \\
Electrons/photons energy scale           			&  $\pm 1$ \\
$b$-tagging                              			&  $\pm 2$ \\
MC statistical uncertainties              			&  $\pm 8$ \\
Other backgrounds normalisation (including $Z+$jets)&  $^{+9}_{-8}$\\
\midrule
Luminosity                         					&  $\pm 2$ \\
\midrule
Total uncertainty                  					& $\pm 26$\\
\noalign{\vskip 1mm}
\hline
\end{tabular}
\end{table}

\section{Phase space for cross-section measurements}
 
\label{sec:phase_space}

The \zgjj electroweak cross-section is measured in a fiducial phase space that is defined to closely follow the selection criteria of the signal region described in Section~\ref{sec:selection}.
This region is defined at the particle level, using stable particles (with a proper decay length $c\tau>10~$mm) which are produced from the hard scattering, including those that are the products of hadronisation, before their interaction with the detector.
Leptons produced in the decay of a hadron, a $\tau$-lepton, or their descendants are not considered in the definition of the fiducial phase space.
Prompt-lepton four-momenta are obtained from a sum of the leptons four-momenta and the four-momenta of photons not originating from hadron decays within a cone of size $\Delta R = 0.1$ around the leptons (`dressed leptons').
Jets are reconstructed using the anti-$k_{t}$ algorithm with radius parameter $ R = 0.4$ using stable particles, excluding electrons, muons, neutrinos, and photons associated with the decay of a $W$ or $Z$ boson.
Photon isolation is defined as the transverse momentum of the system of stable particles within a cone of size $\Delta R = 0.2$ around the photon, excluding muons, neutrinos, and the photon itself.
 
The following selection requirements are imposed to define the fiducial phase space.
The charged leptons from the $Z$ boson decay are required to have transverse momentum $\pt>20~\GeV$ and $\abseta<2.5$, and the invariant mass of the two leptons must be larger than $40~\GeV$.
The photon is required to have transverse momentum $\pt>15~\GeV$ and $\abseta<2.37$. Photons isolated from any hadronic activity are selected by requiring that the photon isolation, as defined above, divided by the photon transverse momentum be less than 5\%.
The angular distance between each of the charged leptons from the $Z$ decay and the photon is required to be $\Delta R>0.4$.
The sum of the two-lepton invariant mass and two-lepton plus photon invariant mass must satisfy ($m_{\ell \ell}+m_{\ell \ell\gamma}$)$>182~\GeV$ in order to exclude final-state radiation events.
 
In addition to these requirements, at least two jets with $\pt>50~\GeV$ and $|\eta|< 4.5$ are required.
The angular distance between each of the charged leptons from the $Z$ boson decay and each of the jets is required to be $\Delta R (j, \ell) >0.3$.
The angular distance between the photon and each of the jets is required to be  $\Delta R (j,\gamma)>0.4$.
The invariant mass $m_{jj}$ of the two highest-\pT jets is required to be $m_{jj}>150~\GeV$, and the difference between the pseudorapidities of these two leading jets is required to be $|\Delta\eta_{jj}|>1$.
Finally, the centrality of the $\ell\ell\gamma$ system relative to the tagging jets must satisfy $\zeta(\ell\ell\gamma)$<5.

\section{Cross-section measurements}
 
\label{sec:result}
 
 
A profile-likelihood-ratio test statistic~\cite{frequentist} is used to measure the signal strength $\mu_{\zgjjew}$ and its associated uncertainties. The systematic uncertainties are treated as nuisance parameters and are constrained with Gaussian distributions in the fit. The statistical uncertainty is determined after having fixed each nuisance parameter to the conditional maximum-likelihood estimator~\cite{frequentist} also called  profiled value. Figure~\ref{fig:BDTScore} shows the BDT score distribution in the signal region, and the $b$-tagged jet multiplicity in the $b$-CR, after having applied the background and signal normalisations and having set the nuisance parameters to the values adjusted by the profile-likelihood fit. The yield of events obtained after the fit is detailed in Table~\ref{tab:VBSyield_postfit}.
 
\begin{figure}[!htbp]
\begin{center}
\subfloat[]{\includegraphics[width=0.45\textwidth]{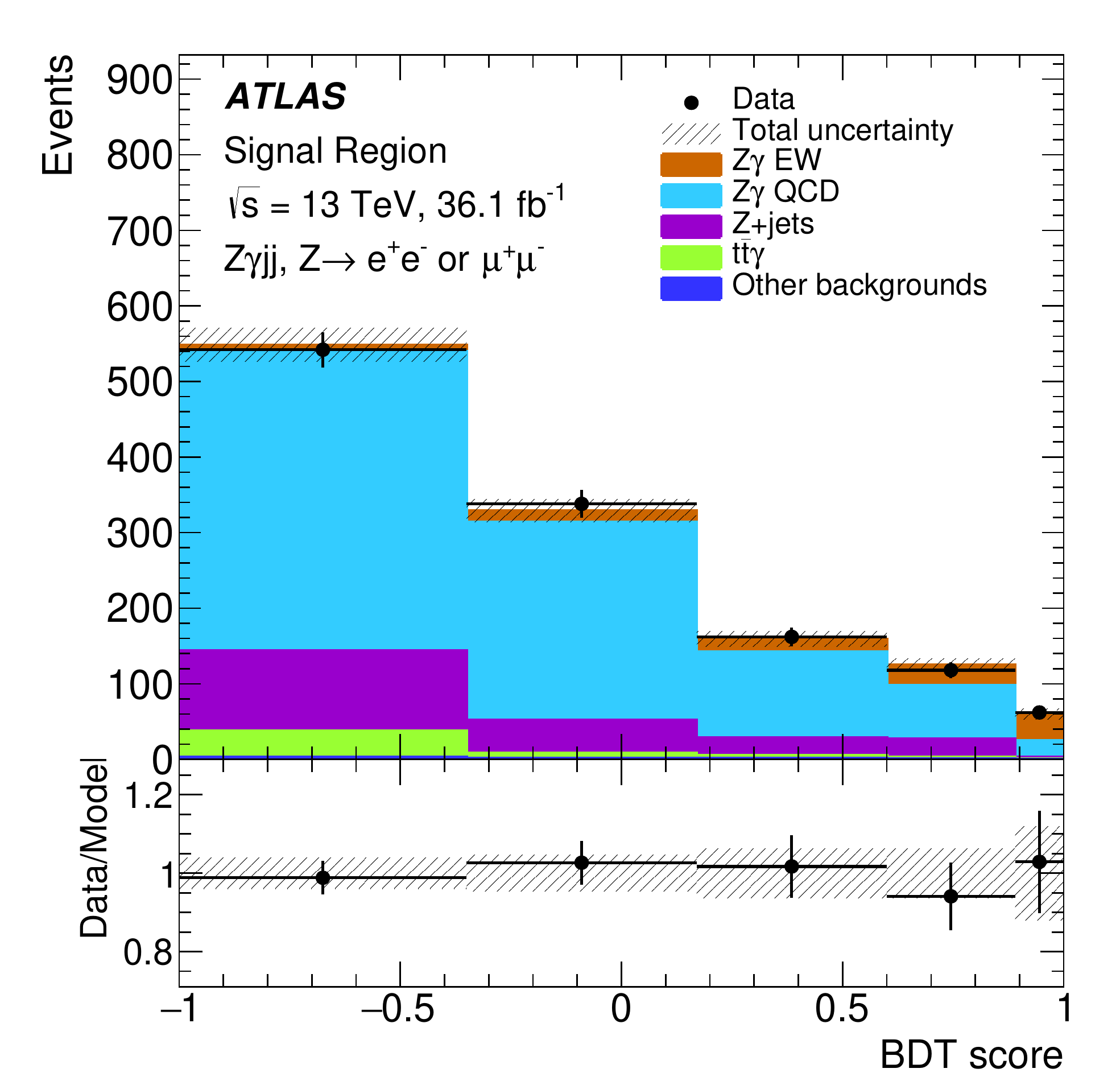}}
\subfloat[]{\includegraphics[width=0.45\textwidth]{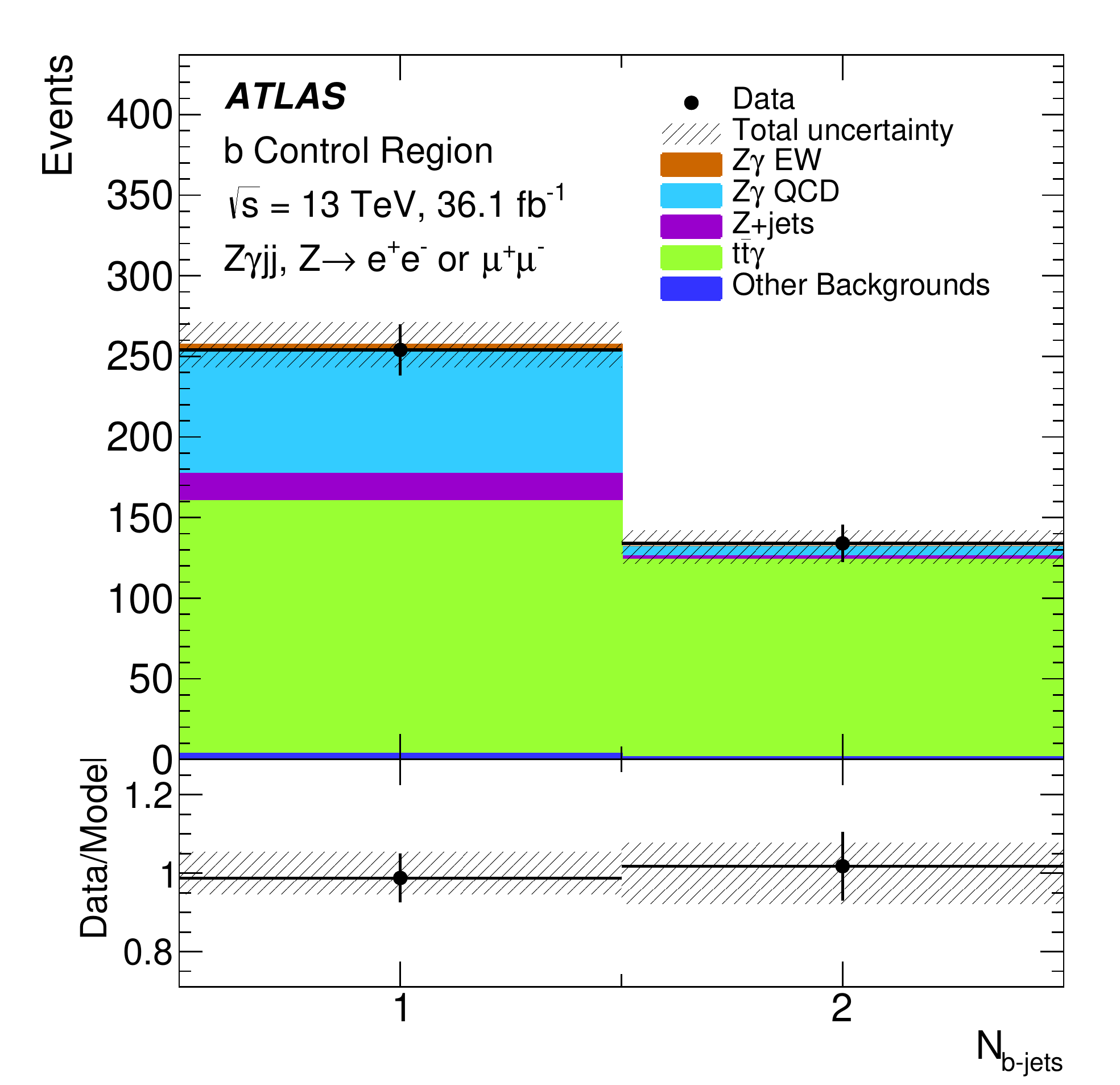}}
\caption{Post-fit distributions of (a) the BDT score in the signal region and of (b) $N_{b\textrm{-jets}}$ in the $b$-CR.
The uncertainty band around the expectation includes all systematic uncertainties and takes into account their correlations as obtained from the fit. Events beyond the upper limit of the histogram are included in the last bin of the $N_{b\textrm{-jets}}$ distribution.}
\label{fig:BDTScore}
\end{center}
\end{figure}

\begin{table}[!htbp] 
\caption{Expected and observed numbers of events in the signal and control regions, after the fit. The expected number of \zgjjew events from \textsc{MadGraph5\_aMC@NLO}~2.3.3 and the estimated number of background events from the other processes are shown. Total post-fit uncertainties, as described in Section~\ref{sec:systematics}, are shown. The uncertainty in the total expected yield is smaller than the quadratic sum of the uncertainties in each process due to these being anti-correlated after the fit.\\
}
\label{tab:VBSyield_postfit}
\centering
\begin{tabular}{
l
S@{$\pm$}S
S@{$\pm$}S
}
\hline
\toprule
&  \multicolumn{2}{c}{SR}    &  \multicolumn{2}{c}{$b$-CR}         \\
\midrule
Data                  & \multicolumn{2}{l}{1222}   & \multicolumn{2}{l}{388}      \\
Total expected       &  1222 & 35                 &  389 &  19                 \\
\midrule
\zgjjew  (signal)     &  104 & 26                  &  5   & 1                \\
\zgjjqcd              &  864 & 60                  &  82  & 9                  \\
$Z+$jets              &  200 & 40                  &  19  & 4                 \\
$t\bar{t}+\gamma $    &  48  & 10                  &  280 & 21                \\
Other backgrounds     &  7   & 1                   &  4   & 1                \\
\bottomrule
\end{tabular}
\end{table}

The signal strength is measured to be:
 
\[
\mu_{\zgjjew} = 1.00 \pm 0.19  \, \mathrm{(stat.)} \pm 0.13 \, \mathrm{(syst.)} ^{+0.13}_{-0.10} \, \mathrm{(mod.)} = 1.00 \pm 0.26 \;, 
\]
 
with ``stat.'' corresponding to the data statistical uncertainty, ``syst.'' to the experimental systematic uncertainties and size of MC samples,
and ``mod.'' to the theoretical modelling of the signal and background MC samples.
This corresponds to a statistical significance of $4.1$ standard deviations both observed and expected.
The expected significance is obtained after having applied the backgrounds normalisations, which are found to be $\mu_{\zgjjqcd} = 0.78^{+0.26}_{-0.20}$ and $\mu_{t\bar{t}+\gamma} = 1.49^{+0.40}_{-0.34}$. The uncertainty in the normalisation parameter values includes the theoretical uncertainties, such as uncertainties derived from QCD scale variations. The normalisation of the $t\bar{t}+\gamma$ background agrees with results from other studies performed with ATLAS Run 2 data~\cite{Aaboud:2018hip} in this final state. The normalisation of the \zgjjqcd background is compatible with results obtained in the $WZjj$ final state~\cite{Aaboud:2018ddq}. It was checked that the signal strength obtained in the combination of the  $ee\gamma{}jj$ and $\mu\mu\gamma{}jj$ final states is consistent with the results obtained separately for each channel.

The fiducial cross-section is measured by computing the product of signal strength and the predicted cross-section used to define it:
\[
\begin{array}{rcl}
\sigma^{\mathrm{fid.}}_{\zgjjew} &=& 7.8 \; \pm 1.5 \,\mathrm{(stat.)} \; \pm 1.0 \,\mathrm{(syst.)} \; ^{+ 1.0} _{- 0.8} \,\mathrm{(mod.)}  \; \mathrm{fb} \, \\
&=& 7.8 \; \pm 2.0 \; \mathrm{fb} .\\
\end{array}
\]
 
This cross-section corresponds to the electroweak \zgjj particle-level cross-section in the fiducial phase space defined in Section~\ref{sec:phase_space} using dressed leptons, and including constructive interference between the signal and the QCD-induced processes.

This measurement can be compared with the LO cross-section predicted by \textsc{MadGraph5\_aMC@NLO}~2.3.3:
\[
\sigma^{\mathrm{fid., \, \textsc{MadGraph}}}_{\zgjjew} = 7.75 \pm 0.03 \,\mathrm{(stat.)} \pm 0.20 \,\mathrm{(PDF+\alpha_\text{S})} \pm 0.40 \,\mathrm{(scale)}  \; \mathrm{fb}.
\]
 
The cross-section predicted by \sherpa~2.2.4 is somewhat larger than that predicted by \textsc{MadGraph5\_aMC@NLO}~2.3.3:
\[
\sigma^{\mathrm{fid., \, \Sherpa}}_{\zgjjew} = 8.94 \pm 0.08 \,\mathrm{(stat.)} \pm 0.20 \,\mathrm{(PDF+\alpha_\text{S})} \pm 0.50 \,\mathrm{(scale)} \; \mathrm{fb} .
\]
 
These predictions do not include the effects of interference, which are estimated with \textsc{MadGraph5\_aMC@NLO}~2.3.3 to be $3\%$ in this phase space.
 
To verify the results of the multivariate analysis, a cut-based approach is also used. The centrality of the $\ell\ell\gamma$ system, $\zeta(\ell\ell\gamma)$, is used as the sensitive variable for the extraction of the electroweak signal. In this alternative method, the signal region is further divided into two regions depending on the dijet invariant mass. The region with $m_{jj}<500~\GeV$ is used to constrain the QCD background, while the region $m_{jj}>500~\GeV$ is kept to extract the signal. The $b$-CR is also kept in this method as a second control region to constrain the $t\bar{t} + \gamma$ background normalisation. The profile-likelihood fit is applied in the same way as described above with the full treatment of systematic uncertainties, allowing exclusion of the background-only hypothesis with a statistical significance of $2.9\sigma$ ($2.7\sigma$) observed (expected). This is compatible with the result obtained with the multivariate approach and with the SM prediction.

The analysis is also used to measure the $Z\gamma jj$ cross-section in the same fiducial phase space. This measurement includes the electroweak-induced signal, the QCD-induced process and their interference. The extraction procedure is the same as for the electroweak \zgjj{} measurement. A template likelihood fit is performed, using the BDT score distribution in the signal region. For this measurement, the $b$-CR is not used, to simplify the statistical model. The omission of this control region in the fit does not change the result. The signal template corresponds to the sum of the \zgjjew and \zgjjqcd templates, the relative amount being fixed to SM expectations. The cross-section is measured to be:
 
\[
\begin{array}{rcl}
\sigma^{\mathrm{fid.}}_{\zgjj} &=& 71 \; \pm 2 \,\mathrm{(stat.)} \; ^{+ 9} _{-7} \,\mathrm{(syst.)} \; ^{+ 21} _{- 17} \,\mathrm{(mod.)}  \; \mathrm{fb} \, \\
&=& 71 \; ^{+ 23} _{- 19} \; \mathrm{fb}.\\
\end{array}
\]

This agrees with the SM predictions obtained by summing the \zgjjew and \zgjjqcd predictions obtained from \textsc{MadGraph5\_aMC@NLO}~2.3.3 and \sherpa~2.2.2, respectively,
 
\[
\sigma^{\mathrm{fid., \, \textsc{MadGraph}+\Sherpa}}_{\zgjj} = 88.4 \pm 2.4 \,\mathrm{(stat.)} \pm 2.3 \,\mathrm{(PDF+\alpha_\text{S})} ^{+29.4}_{-19.1} \,\mathrm{(scale)}  \; \mathrm{fb} .
\]

\FloatBarrier

\section{Conclusion}

Evidence for electroweak production of two jets in association with a \zg pair is presented using $36.1~\ifb$ of $pp$ collision data at $\sqrt{s} = 13$~\TeV\ collected with the ATLAS detector at the LHC. The production cross-section of this process is measured in a fiducial phase space approximating the acceptance of the analysis. This measurement uses the leptonic decay of the $Z$ boson into $e^{+}e^{-}$ or $\mu^{+}\mu^{-}$. The measurement is performed using a BDT to enhance the signal to background-ratio. The dominant backgrounds, $\zgjjqcd$, $Z+$jets and $t\bar{t}+\gamma $  are all estimated from the data.
The background-only hypothesis is excluded with observed and expected significances of $4.1$ standard deviations.
 
In the fiducial phase space the electroweak cross-section of the $Z\gamma jj$ process is measured to be:
\[
\begin{array}{rcl}
\sigma^{\mathrm{fid.}}_{\zgjjew} &=& 7.8 \; \pm 1.5 \,\mathrm{(stat.)} \; \pm 1.0 \,\mathrm{(syst.)} \; ^{+ 1.0} _{- 0.8} \,\mathrm{(mod.)} \; \mathrm{fb}, \\
&=& 7.8 \; \pm 2.0 \; \mathrm{fb} \\
\end{array}
\]
 
in good agreement with the Standard Model predictions at \LO in perturbative QCD.

\section*{Acknowledgements}
 

We thank CERN for the very successful operation of the LHC, as well as the
support staff from our institutions without whom ATLAS could not be
operated efficiently.
 
We acknowledge the support of ANPCyT, Argentina; YerPhI, Armenia; ARC, Australia; BMWFW and FWF, Austria; ANAS, Azerbaijan; SSTC, Belarus; CNPq and FAPESP, Brazil; NSERC, NRC and CFI, Canada; CERN; CONICYT, Chile; CAS, MOST and NSFC, China; COLCIENCIAS, Colombia; MSMT CR, MPO CR and VSC CR, Czech Republic; DNRF and DNSRC, Denmark; IN2P3-CNRS and CEA-DRF/IRFU, France; SRNSFG, Georgia; BMBF, HGF and MPG, Germany; GSRT, Greece; RGC and Hong Kong SAR, China; ISF and Benoziyo Center, Israel; INFN, Italy; MEXT and JSPS, Japan; CNRST, Morocco; NWO, Netherlands; RCN, Norway; MNiSW and NCN, Poland; FCT, Portugal; MNE/IFA, Romania; MES of Russia and NRC KI, Russia Federation; JINR; MESTD, Serbia; MSSR, Slovakia; ARRS and MIZ\v{S}, Slovenia; DST/NRF, South Africa; MINECO, Spain; SRC and Wallenberg Foundation, Sweden; SERI, SNSF and Cantons of Bern and Geneva, Switzerland; MOST, Taiwan; TAEK, Turkey; STFC, United Kingdom; DOE and NSF, United States of America. In addition, individual groups and members have received support from BCKDF, CANARIE, Compute Canada and CRC, Canada; ERC, ERDF, Horizon 2020, Marie Sk{\l}odowska-Curie Actions and COST, European Union; Investissements d'Avenir Labex, Investissements d'Avenir Idex and ANR, France; DFG and AvH Foundation, Germany; Herakleitos, Thales and Aristeia programmes co-financed by EU-ESF and the Greek NSRF, Greece; BSF-NSF and GIF, Israel; CERCA Programme Generalitat de Catalunya and PROMETEO Programme Generalitat Valenciana, Spain; G\"{o}ran Gustafssons Stiftelse, Sweden; The Royal Society and Leverhulme Trust, United Kingdom.
 
The crucial computing support from all WLCG partners is acknowledged gratefully, in particular from CERN, the ATLAS Tier-1 facilities at TRIUMF (Canada), NDGF (Denmark, Norway, Sweden), CC-IN2P3 (France), KIT/GridKA (Germany), INFN-CNAF (Italy), NL-T1 (Netherlands), PIC (Spain), ASGC (Taiwan), RAL (UK) and BNL (USA), the Tier-2 facilities worldwide and large non-WLCG resource providers. Major contributors of computing resources are listed in Ref.~\cite{ATL-GEN-PUB-2016-002}.
 


\printbibliography

\clearpage 
 
\begin{flushleft}
\hypersetup{urlcolor=black}
{\Large The ATLAS Collaboration}

\bigskip

\AtlasOrcid[0000-0002-6665-4934]{G.~Aad}$^\textrm{\scriptsize 102}$,    
\AtlasOrcid[0000-0002-5888-2734]{B.~Abbott}$^\textrm{\scriptsize 129}$,    
\AtlasOrcid{D.C.~Abbott}$^\textrm{\scriptsize 103}$,    
\AtlasOrcid[0000-0002-2788-3822]{A.~Abed~Abud}$^\textrm{\scriptsize 71a,71b}$,    
\AtlasOrcid{K.~Abeling}$^\textrm{\scriptsize 53}$,    
\AtlasOrcid{D.K.~Abhayasinghe}$^\textrm{\scriptsize 94}$,    
\AtlasOrcid[0000-0002-8496-9294]{S.H.~Abidi}$^\textrm{\scriptsize 167}$,    
\AtlasOrcid[0000-0002-8279-9324]{O.S.~AbouZeid}$^\textrm{\scriptsize 40}$,    
\AtlasOrcid{N.L.~Abraham}$^\textrm{\scriptsize 156}$,    
\AtlasOrcid[0000-0001-5329-6640]{H.~Abramowicz}$^\textrm{\scriptsize 161}$
% ...
\bigskip
\\

% ...
$^{40}$Niels Bohr Institute, University of Copenhagen, Copenhagen; Denmark.\\
% ...
$^{53}$II. Physikalisches Institut, Georg-August-Universit\"{a}t G\"ottingen, G\"ottingen; Germany.\\
% ...
$^{71}$$^{(a)}$INFN Sezione di Pavia;$^{(b)}$Dipartimento di Fisica, Universit\`a di Pavia, Pavia; Italy.\\
% ...
$^{94}$Department of Physics, Royal Holloway University of London, Egham; United Kingdom.\\
% ...
$^{102}$CPPM, Aix-Marseille Universit\'e, CNRS/IN2P3, Marseille; France.\\
$^{103}$Department of Physics, University of Massachusetts, Amherst MA; United States of America.\\
% ...
$^{129}$Homer L. Dodge Department of Physics and Astronomy, University of Oklahoma, Norman OK; United States of America.\\
% ...
$^{156}$Department of Physics and Astronomy, University of Sussex, Brighton; United Kingdom.\\
% ...
$^{161}$Raymond and Beverly Sackler School of Physics and Astronomy, Tel Aviv University, Tel Aviv; Israel.\\
% ...
$^{167}$Department of Physics, University of Toronto, Toronto ON; Canada.\\
% ...
$^{a}$ Also at Borough of Manhattan Community College, City University of New York, New York NY; United States of America.\\
$^{b}$ Also at CERN, Geneva; Switzerland.\\
% ...

\end{flushleft}


\end{document}